\documentclass[11pt,twoside,a4paper]{article}
\usepackage{myartmods}
\usepackage[utf8]{inputenc}
\usepackage{datetime}\usdate
\usepackage[english]{babel}

\usepackage{amsmath}
\usepackage{amsfonts} 
\usepackage{amssymb}
\usepackage{mathtools}
\usepackage{multirow,bigdelim}
\usepackage{gensymb}

\usepackage[sort]{cite}
\usepackage{graphicx}
\usepackage{caption}
\usepackage{subcaption}
\usepackage[bottom]{footmisc}
\usepackage{tabularx}
\usepackage{booktabs} 
\usepackage{authblk}

\numberwithin{equation}{section}
\numberwithin{figure}{section}
\numberwithin{table}{section}

\catcode`@=11
\def\minim{\mathop{\operator@font minimize}}
\def\minimize#1{{\displaystyle\minim_{#1}}}

\def\subject{\mathop{\operator@font subject\ to}}
\catcode`@=12

\begin{document}
\title      {On tradeoffs between treatment time and plan quality of volumetric-modulated arc therapy with sliding-window delivery}
\author[1,2]{Lovisa Engberg\thanks{Corresponding author: loven140@kth.se}}
\author[1]  {Anders Forsgren}
\affil[1]   {Optimization and Systems Theory, Department of Mathematics, \protect\\KTH Royal Institute of 
             Technology, Stockholm SE-100 44, Sweden}
\affil[2]   {RaySearch Laboratories, Sveav\"{a}gen 44, Stockholm SE-103 65, Sweden}
\date       {Manuscript\\\today}
\markboth   {On tradeoffs between treatment time and plan quality of sliding-window VMAT}{On tradeoffs between treatment time and plan quality of sliding-window VMAT}
\maketitle\thispagestyle{empty}

\begin{abstract} 
The purpose of this study is to give an exact formulation of optimization of volumetric-modulated arc therapy (VMAT) with sliding-window delivery, and to investigate the plan quality effects of decreasing the number of sliding-window sweeps made on the 360-degree arc for a faster VMAT treatment. In light of the exact formulation, we interpret an algorithm previously suggested in the literature as a heuristic method for solving this optimization problem. By first making a generalization, we suggest a modification of this algorithm for better handling of plans with fewer sweeps. In a numerical study involving one prostate and one lung case, plans with varying treatment times and number of sweeps are generated. It is observed that, as the treatment time restrictions become tighter, fewer sweeps may lead to better plan quality. Performance of the original and the modified version of the algorithm is evaluated in parallel. Applying the modified version results in better objective function values and less dose discrepancies between optimized and accurate dose, and the advantages are pronounced with decreasing number of sweeps. 

\end{abstract}

\section{Introduction}
A clinical advantage of volumetric-modulated arc therapy (VMAT) over static-gantry delivery of radiotherapy, is the potential to obtain a shortened treatment time without compromising plan quality. In static delivery, the treatment is limited to a few angles around the patient, and the beam is turned off while the gantry moves to the next angle. VMAT delivery, on the other hand, allows the gantry to rotate during irradiation and multileaf collimation. Treatment time savings are achieved since the beam is never turned off. 

From a treatment planning perspective, a delivery technique and its clinical advantages can be implemented first when there is a way to mathematically formulate and efficiently solve (at least approximately) the associated optimization problem, so that a planning tool can eventually be developed. VMAT is clinical routine since more than a decade thanks to dedicated research and development reflected in the many approaches to VMAT treatment planning proposed in the literature. As noticed by Peng et al.~\cite{peng2012}, the literature is mainly focused on algorithms of a heuristic nature---possibly with some local-optimizing steps given a restricted optimization formulation, but seldom related to a complete VMAT optimization formulation---due to the mathematical complexity associated with the continuously rotating gantry. Suggested approaches are often divided into two categories. In two-phase algorithms, idealized fluence profiles at certain gantry angles are first generated by solving a fluence map optimization (FMO) problem. Both coarse \cite{bzdusek2009} and dense angular discretizations \cite{craft2012a} have been used. The optimized fluence profiles are then transformed into leaf trajectories in an arc-sequencing phase, where all delivery constraints are taken into account. The objective of arc-sequencing varies: algorithms and/or formulations have been suggested that either amount to find the leaf trajectories that best reproduce the optimal fluence profile (see, e.g., \cite{shepard2007,craft2012a}) or that best replicate the delivered target dose (see, e.g., \cite{wang2008}). The other approach to VMAT optimization is to directly take the deliverability of the treatment plan into account by (approximately) solving a so-called direct machine parameter optimization (DMPO) problem. DMPO formulations of VMAT delivery are in general nonconvex and considered more complex than the static-gantry counterparts. On the other hand, as reported by Shepard et al.~\cite{shepard2002} and Rao et al.~\cite{rao2010}, targeting a DMPO formulation often results in improved plan quality for both static-gantry and VMAT delivery as compared to applying two-phase algorithms, and can be motivated in that aspect. Bzdusek et al.~\cite{bzdusek2009} describe an efficient method that combines a two-phase algorithm and a nonconvex DMPO formulation, the former giving the initial solution to a gradient-based method for solving the latter to local optimality. The method has been adopted by several commercial systems for VMAT treatment planning \cite{unkelbach2015}. Peng et al.~\cite{peng2012,peng2015} suggest a column-generation and a heuristic decomposition approach to handle a fully stated DMPO formulation. Classical heuristic methods have also been applied, including simulated annealing \cite{otto2007} and tabu search \cite{ulrich2007}. A comprehensive review of approaches to VMAT optimization is given by Unkelbach et al. \cite{unkelbach2015}. 

In Papp and Unkelbach~\cite{papp2014}, an algorithm is presented to handle DMPO for VMAT delivery restricted to unidirectional leaf trajectories---``sliding windows'' or sweeps. By considering as variables the times of arrival and departure of the leaves at fixed positions (bixels) along the sweeping direction, the authors demonstrate that the set of sweeps can be expressed using linear inequalities, and that the resultant radiation fluence passing through the bixels is given by a linear, hence convex, function of the arrival and departure times. This opportunity does not occur for regular (with arbitrary leaf motions) VMAT delivery, which requires nonconvex formulations to exactly model the fluence profiles as a function of leaf positions \cite{unkelbach2015}. Unfortunately, nonconvexities eventually catch also sliding-window VMAT, since the computation of dose is a nonconvex operation due to the rotation of the gantry. The algorithm suggested by Papp and Unkelbach therefore amounts to solving a sequence of simplified DMPO formulations (subproblems) with approximate linear dose computations; accurate dose is computed first as a final step. In the present study, we express the accurate dose as an explicit function of the sliding window sweeps to obtain a full DMPO formulation. The purpose is to formalize sliding-window VMAT optimization for further development of algorithms. In particular, in light of the suggested exact formulation of sliding-window VMAT optimization, we interpret the Papp and Unkelbach algorithm as a heuristic method for solving this optimization problem and suggest a generalization of the subproblem update scheme. 

Except for the linear representation of leaf trajectories and resultant fluence, another notable benefit of sliding-window over regular VMAT is the opportunity to create a new deliverable plan by convex combination of other sliding-window VMAT plans~\cite{craft2014}. This property is particularly interesting for multicriteria optimization, as it enables Pareto set navigation in the domain of deliverable plans. It also motivates further development of methods to handle the associated optimization problems. 

A drawback with the sliding-window approach is that the many unidirectional sweeps back and forth over the fluence field increase the treatment time as compared to regular VMAT delivery, especially for cases with large targets and thus wide fields to traverse~\cite{unkelbach2015}. In the studies by Papp and Unkelbach~\cite{papp2014} and Craft et al.~\cite{craft2012a}, delivery times of 3-6 minutes are needed to achieve the desired plan quality, whereas 1-2 minutes are expected for regular VMAT~\cite{bzdusek2009}. The criticism regarding delivery time is justified since, again, a shortened treatment was one of the main arguments for choosing VMAT over static delivery in the first place. Therefore, in our numerical study, we investigate the effect on plan quality of decreasing the number of sweeps made on the 360-degree arc for a faster treatment. Besides explicitly limiting the treatment time during optimization, controlling the number of sweeps is a potential means to trade plan quality for a more efficient VMAT delivery, analogous to limiting the number of beams in static-gantry delivery. To generate plans, we suggest and apply a version of the generalization of the Papp and Unkelbach algorithm. Our suggested version is designed to better handle a setting with fewer sweeps delivered on relatively large portions (arc segments) of the 360-degree arc.

\section{Method}\label{sec:Methods}
\subsection{Optimization formulation}
The planning objectives and constraints used in this study follow the formulation suggested in our previous works~\cite{engberg2017,engberg2018}:
\begin{equation}\label{eq:propForm}
\begin{aligned}
& \minimize{d,\,\xi} 
           && \mkern-10mu \mathmakebox[0pt][l]{\big[\,\xi_1, \cdots, \xi_q, -\xi_{q+1}, \cdots, -\xi_K\,\big]^T} \\
& \subject && D^+(d;v_k,s_k) \leq \xi_k \leq u_k, && \hat{l}_k \leq \xi_k, && k = 1,\ldots,q,       \\ 
&          && D^-(d;v_k,s_k) \geq \xi_k \geq l_k, && \hat{u}_k \geq \xi_k, && k = q\!+\!1,\ldots,K, \\
&          && \mathmakebox[0pt][l]{d \text{ deliverable dose distribution,}}
\end{aligned}
\end{equation} 
where $D^+(\cdot\,;v,s)$ and $D^-(\cdot\,;v,s)$ denote the upper and lower mean-tail-dose functions for volume fraction $v$ in structure $s$. The mean-tail-dose functions were introduced by Romeijn et al.~\cite{romeijn2006} and are in our research used as convex approximations of the dose-at-volume function frequently used to evaluate plan quality. All dose constraints of \eqref{eq:propForm} can be expressed by a set of linear inequalities (see Appendix A of our previous work~\cite{engberg2018} for a fully expanded formulation). Depending on the deliverability constraints, the optimization problem can therefore be a linear or convex program, or a general non-convex program.

\subsection{Accurate dose computation for sliding-window VMAT}
The computation of accurate dose is added as a final step in the sliding-window VMAT optimization algorithm by Papp and Unkelbach~\cite{papp2014}, but not stated as an explicit function. In this section, we formulate the accurate dose as a function of deliverable leaf trajectories, and show how this nonsmooth function can be incorporated into a mixed integer linear program (MILP) to form an exact formulation of sliding-window VMAT optimization. 

Let $b$, $b=1,\ldots,B$, enumerate the arc segments of the 360-degree arc, within each one sweep is to be delivered. Let the sweeping trajectories of the leaves be represented as in \cite{papp2014} and denoted as in \cite{engberg2018}, i.e., by the points in time $r_{b,n,j}$ and $l_{b,n,j}$ when respectively the leading and the trailing leaf of leaf pair $n$, $n=1,\ldots,N$, when regarded in the $b$th arc segment, begin traversing bixel $j$, $j=1,\ldots,J$. Furthermore, let $k$, $k=1,\ldots,K$, enumerate the control points on the 360-degree arc, equiangularly distributed at angles $k\theta$ (typical control point spacings are $\theta = 2\degree$ or $4\degree$). Each control point angle $k\theta$ has an associated dose deposition matrix $P^{k\theta}$ which, in accordance with the final accurate dose computation in \cite{papp2014}, is assumed valid in the $\theta$-neighborhood $\left[(k-\frac{1}{2})\theta,(k+\frac{1}{2})\theta\right]$. 

\begin{figure}\centering
	\includegraphics[scale=.75]{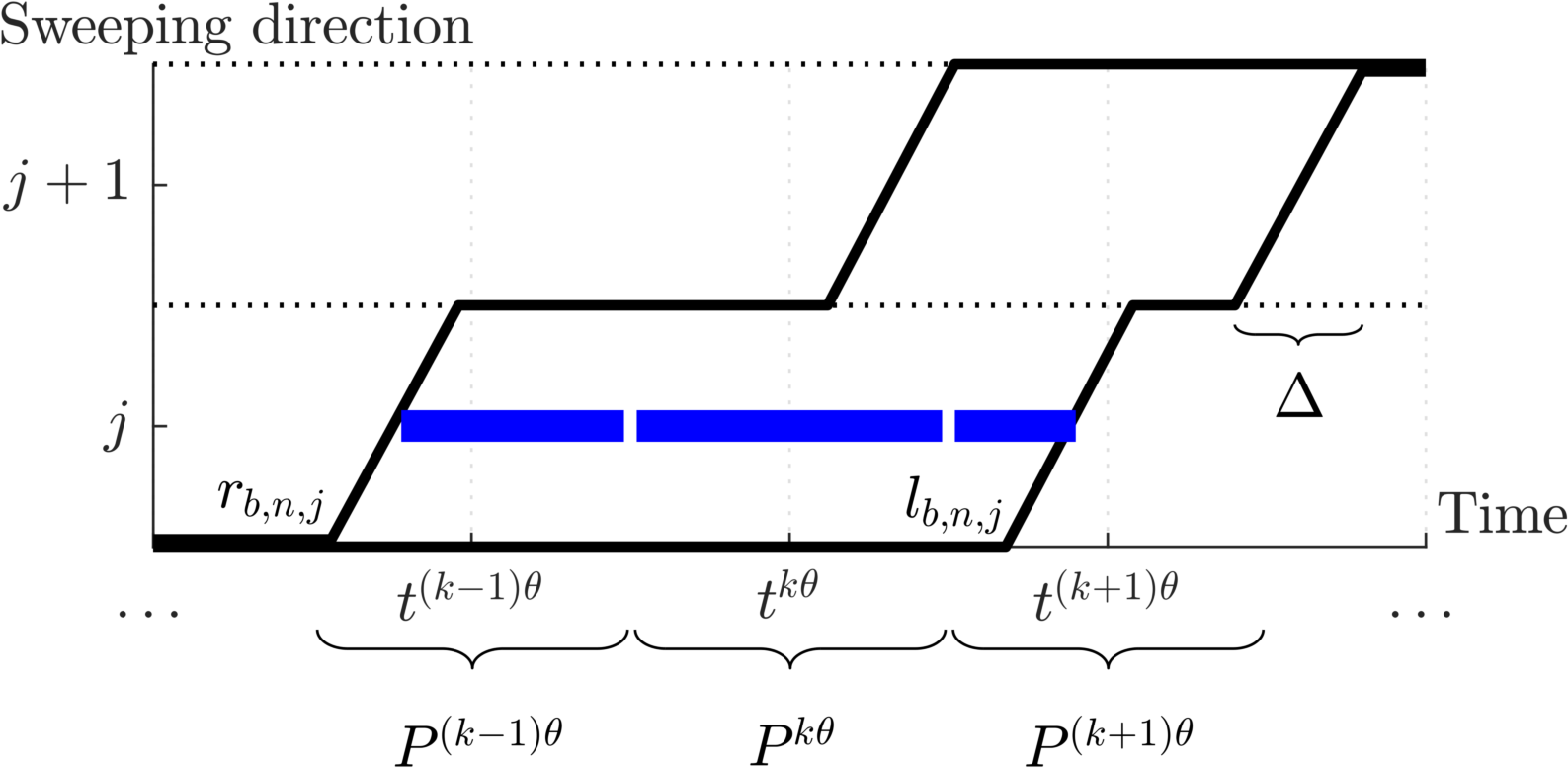}
	\caption{A sweep example. The leaf trajectory is represented by the points in time $r_{b,n,j}$ and $l_{b,n,j}$ when respectively the leading and the trailing leaf begin traversing bixel $(b,n,j)$, with $\Delta$ the constant traversing time (see text for further definitions). The blue ribbons mark the exposure of bixel $(b,n,j)$ at control points $k-1$, $k$, and $k+1$.}
	\label{fig:leaftrajectory}
\end{figure}

As the gantry rotates across the arc segments, a number of control points will be traversed. Let $K_b$, $K_b \subset \{1,\ldots,K\}$, denote the set that collects the consecutive control points passed by arc segment $b$. The rotating speed of the gantry is assumed constant over each arc segment, and is determined by the time when all leaves have finished traversing the field; let $t^{k\theta}$ denote the point in time when the gantry passes angle $k\theta$ (in turn given by the speed of the gantry). To formulate the dose distribution as a function of leaf trajectories, we first note that the exposure of bixel $(b,n,j)$ at control point $k \in K_b$, illustrated in the trajectory plot of Figure~\ref{fig:leaftrajectory}, equals the quantity 
\[
	\max\big\{\min\big(l_{b,n,j}+\frac{\Delta}{2},\,t^{(k+\frac{1}{2})\theta}\big) - \max\big(r_{b,n,j}+\frac{\Delta}{2},\,t^{(k-\frac{1}{2})\theta}\big),\, 0\, \big\},
\]
where $\Delta$ is the constant bixel traversing time (assuming, e.g., that the leaves are always travelling with maximum speed while in motion). In this expression, the inner $\min$ and $\max$ functions localize the beginning and end of the exposure, respectively, and the outer $\max$ function transforms any negative value to zero exposure. 
The dose in voxel $i$ is given by the sum of exposure contributions from all bixels, and can thus be written 
\begin{multline}\label{eq:doseComp}
		d_i = \delta \sum_{b,n,j} \sum_{k \in K_b} P^{k\theta}_{i,(b,n,j)}\, \max\big\{\min\big(l_{b,n,j}+\frac{\Delta}{2},\,t^{(k+\frac{1}{2})\theta}\big)\, - \\ - \max\big(r_{b,n,j}+\frac{\Delta}{2},\,t^{(k-\frac{1}{2})\theta}\big),\, 0\, \big\},
\end{multline}
where $\delta$ is the constant dose rate.\footnote{Note that $r_{b,n,j}$ and $l_{b,n,j}$ alternate between denoting the traversing time of the right and left leaf, depending on the sweeping direction (left-to-right or right-to-left) used in arc segment $b$.} Since involving the nonsmooth $\min$ and $\max$ functions, it is not clear how to handle the accurate dose computation in \eqref{eq:doseComp} in a smooth optimization setting. It is possible, however, to transform all dose constraints of \eqref{eq:propForm} into a MILP formulation by the introduction of artificial and binary variables; its derivation is delayed to Appendix~\ref{app:MILP} as no attempt is made in this study to solve the exact MILP formulation. The MILP formulation requires the introduction of six binary variables per bixel and control point, adding up to the order of $10^5$ binary variables. While the sequential nature of the unidirectional leaf sweeps likely allows the construction of several types of valid inequalities, which has not been investigated in this study, we envisage that applying standard methods to solve the MILP formulation to proven optimality would be too time-consuming for our application. It should also be noted that the MILP formulation requires the use of the ``big $M$'' method where a large-valued parameter $M$ is included, which is known to be prone to numerical instability in combination with standard methods. 

In the following, for readability, we consider a renumbering of the bixels that enables enumeration by one index instead of three. We reuse $j$ as the new index since its association with a bixel has already been established (i.e., $j=1,\ldots,BNJ$ in the following). Furthermore, for ease of notation, we will utilize the fact that $j$ determines the arc segment membership, thus the set $K_b$, given the chosen renumbering mapping. 

\subsection{Heuristic methods}
We refer to the optimization formulation \eqref{eq:propForm} with the accurate dose computation given in \eqref{eq:doseComp} as the exact problem. 

Given the exact problem, we are in a position to interpret the algorithm by Papp and Unkelbach~\cite{papp2014} as a heuristic method to find an approximate solution. Its heuristic nature is due to the approximate dose computation used during optimization. In the algorithm, the dose deposition \emph{column} for bixel $j$ is assumed constant over the arc segment, hence does not change as the gantry passes new control points. The fixed column, here denoted by $P_j$, is chosen among all the available columns $P^{k\theta}_j, k \in K_b$. The effect is a linear but approximate dose computation,  
\[
	d_i = \delta \sum_j P_j (l_j-r_j). 
\] 
The algorithm amounts to solving the resulting smooth subproblem repeated times with updated choices of $P_j$ depending on the previous solution. More specifically, $P^{k\theta}_j$ of the control point whose $\theta$-neighborhood contains the midpoint of the exposure interval for bixel $j$ is chosen; $P^{k\theta}_j$ of the control point closest to the midpoint of the arc segment is chosen initially. The method terminates when the new choice is sufficiently similar to the previous one. 

Now, in terms of the exact problem, the Papp and Unkelbach algorithm solves a sequence of relaxations of restrictions. To see this, we introduce variables $p_j^k$ to denote the \emph{fraction} of the total exposure of bixel $j$ occurring at control point $k \in K_b$, 
\begin{equation}\label{eq:exposureFraction}
	p_j^k = \max\big\{\min\big(l_j+\frac{\Delta}{2},\,t^{(k+\frac{1}{2})\theta}\big) - \max\big(r_j+\frac{\Delta}{2},\,t^{(k-\frac{1}{2})\theta}\big),\, 0\, \big\}\, \frac{1}{l_j-r_j}, 
\end{equation}
so that the exact dose can be written
\begin{equation}\label{eq:doseComp2}
	d_i = \sum_j \Big( \sum_{k \in K_b} P^{k\theta}_{ji} p_j^k \Big)\, (l_j-r_j).
\end{equation}
By construction, we have that $\sum_{k \in K_b} p_j^k = 1$ and $p_j^k \in [0,1]$ which implies that the column $\sum_{k \in K_b} P^{k\theta}_j p_j^k$ is a convex combination of the columns $P^{k\theta}_j$. Constructing a subproblem of the algorithm is equivalent to treating the $p_j^k$'s as parameters and fixing them to binary values: for each bixel, $p_j^k = 1$ for the control point for which $P_j$ was chosen and $p_j^k = 0$ for the others. The interpretation of this maneuver is that the sweeps are assumed resulting in bixel exposures concentrated to one pre-determined control point. Besides the fixation of $p_j^k$, a proper restriction of the exact problem along this assumption needs the additional bounds
\begin{equation}\label{eq:boundsForRestriction}
	r_j + \frac{\Delta}{2} \geq t^{(k_j-\frac{1}{2})\theta} \quad\text{and}\quad
	l_j + \frac{\Delta}{2} \leq t^{(k_j+\frac{1}{2})\theta}, 
\end{equation}
denoting by $k_j$ the control point $k$ for which $p_j^k = 1$. However, the bounds are not included in the subproblems, which thus may be interpreted as relaxations of this restriction. It should be noted that the restricted problem is of little interest in practice, since the underlying assumption of concentrated bixel exposure is highly conservative. 

In light of the exact problem, we are also in a position to make a generalization of the Papp and Unkelbach algorithm. A generalization is obtained by allowing any fractional values when updating the fixed $p_j^k$'s. As with binary values, fractional values of $p_j^k$ give a linear approximate dose computation favorable for optimization; but also have the ability to reflect bixel exposures that span several consecutive control points within the arc segment, thus have the potential to give a better approximation of accurate dose. A benefit of using a high-quality approximate dose computation during the optimization process is less deterioration of the optimal solution after accurate dose has been computed. While only minor such dose deviations are reported in the numerical study of the original paper~\cite{papp2014}, the explanation given is the observation that the exposure of most bixels of the 18-degree arc segments lasts no more than one control point. For longer arc segments (i.e., fewer sweeps) with more control points to pass, it is likely that a similar observation can no longer be made. We therefore suggest a modification of the Papp and Unkelbach algorithm, where the $p_j^k$ are assigned the exact exposure fraction in \eqref{eq:exposureFraction} obtained for the previous solution. The sought-after benefit is better handling of long arc segments. 

A note regarding convergence is needed. The Papp and Unkelbach algorithm is terminated once the updated values of the $p_j^k$'s are sufficiently close (in a given metric) to the previous values. There is thus an expectation, supported by the numerical results of \cite{papp2014}, that the algorithm reaches a state with only slight changes in updates. Unfortunately from a mathematical perspective, convergence of the algorithm in this sense cannot be related to the globally optimal solution of the exact problem; nor can it be guaranteed that the algorithm produces a monotonically improving sequence of solutions. The convergence situation does not change with our suggested modification, i.e., with $p_j^k$ set to the exact exposure fraction.

\section{Results}\label{sec:Results}
The effects of decreasing the number of sweeps in order to obtain a faster treatment is studied for one prostate and one lung case. The optimized plan quality in terms of objective function value is evaluated, as well as the performance of the suggested fractional version of the generalization of the Papp and Unkelbach algorithm compared to the original binary version. For simplicity of notation, the two versions of the algorithm are henceforth referred to as the \emph{fractional} and the \emph{binary} version, respectively. 

All dose deposition matrices and patient data is exported from RayStation (RaySearch Laboratories, Stockholm, Sweden) to MATLAB. We consider a scalarized weighted-sum instance of the multicriteria formulation in \eqref{eq:propForm} constructed by accumulating the objective functions using positive weighting factors. Combined with the linear sliding-window deliverability constraints and the linear approximate dose computation, the subproblems of the two algorithm versions become linear programs. A tailored interior-point method implemented in MATLAB that exploits the structure of these linear programs is used to solve the sequence of subproblems; we refer to our previous work~\cite{engberg2018} for a description of the interior-point method. After each subproblem solve, accurate dose is computed for performance analysis purposes.

Treatment plans of different treatment time restrictions and number of sweeps are generated using both the fractional and the binary version of the algorithm; delivery of 7, 11, and 20 sweeps in a maximum time of 240, 180, and 120 seconds are considered for both patient cases (120 seconds relaxed to 150 seconds for the 20-sweep plans due to infeasibility). A 4-degree control point spacing is used. The rotating speed of the gantry is limited to between 4.8 and 0.5 degrees per second, implying a minimum of 75 seconds to rotate through the 360-degree arc.  

\begin{figure}
	\centering
	\begin{subfigure}[b]{.7\textwidth}
		\flushleft\hskip7pt
		\hskip6pt\includegraphics[scale=.59]{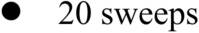}\hskip20pt
		\hskip13pt\includegraphics[scale=.59]{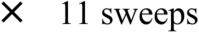}\hskip20pt
		\hskip13pt\includegraphics[scale=.59]{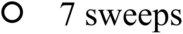}
	\end{subfigure}
	\vskip-.25\baselineskip
	\begin{subfigure}[b]{.7\textwidth}
		\flushleft\hskip7pt
		\includegraphics[scale=.59]{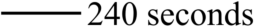}\hskip20pt
		\includegraphics[scale=.59]{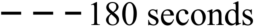}\hskip20pt
		\raisebox{-1.5pt}{\includegraphics[scale=.59]{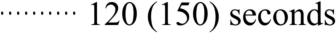}}
	\end{subfigure}
	\vskip\baselineskip
	\begin{subfigure}[b]{.32\textwidth}
		\centering
		\includegraphics[scale=.75]{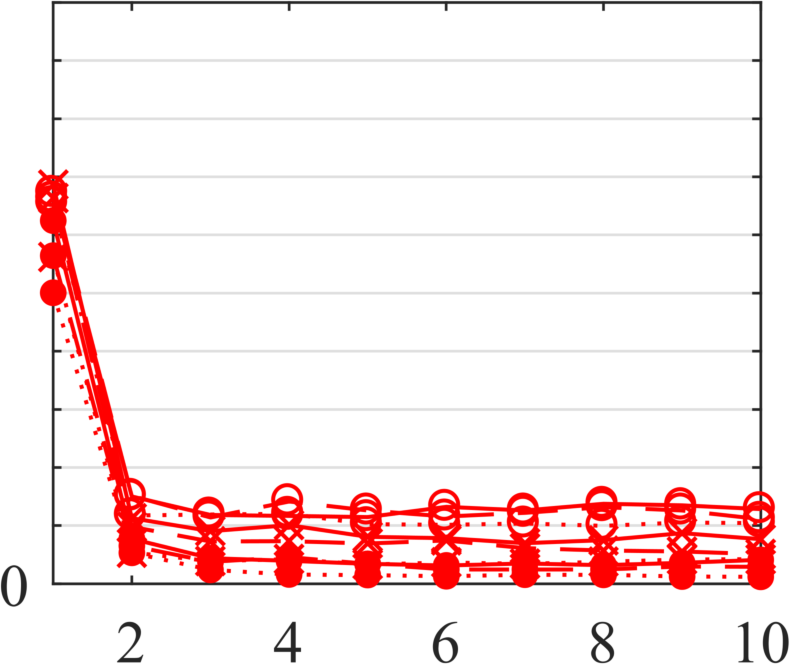}
	\end{subfigure}
	\hskip3pt
	\begin{subfigure}[b]{.32\textwidth}
		\centering
		\includegraphics[scale=.75]{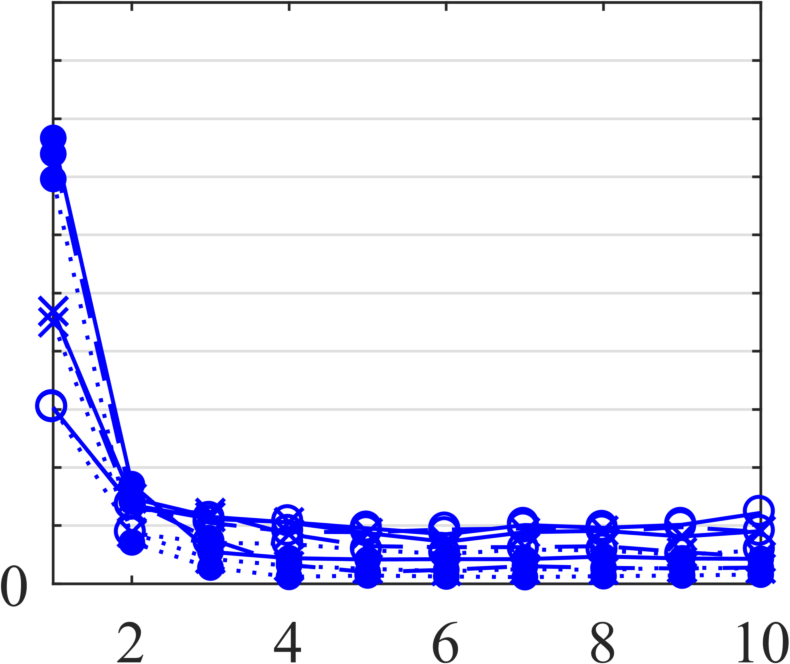}
	\end{subfigure}
	\vskip\baselineskip
	\begin{subfigure}[b]{.32\textwidth}
		\centering
		\includegraphics[scale=.75]{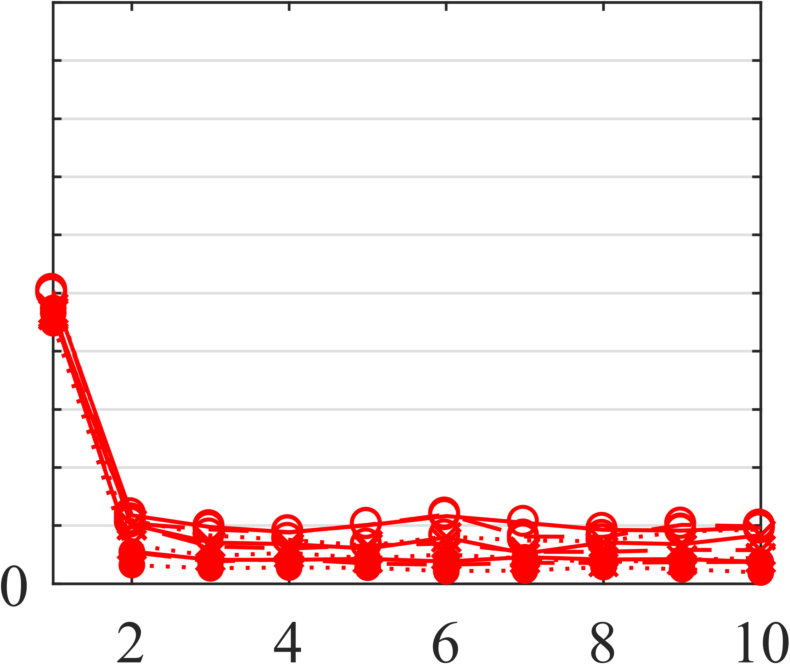}
	\end{subfigure}
	\hskip3pt
	\begin{subfigure}[b]{.32\textwidth}
		\centering
		\includegraphics[scale=.75]{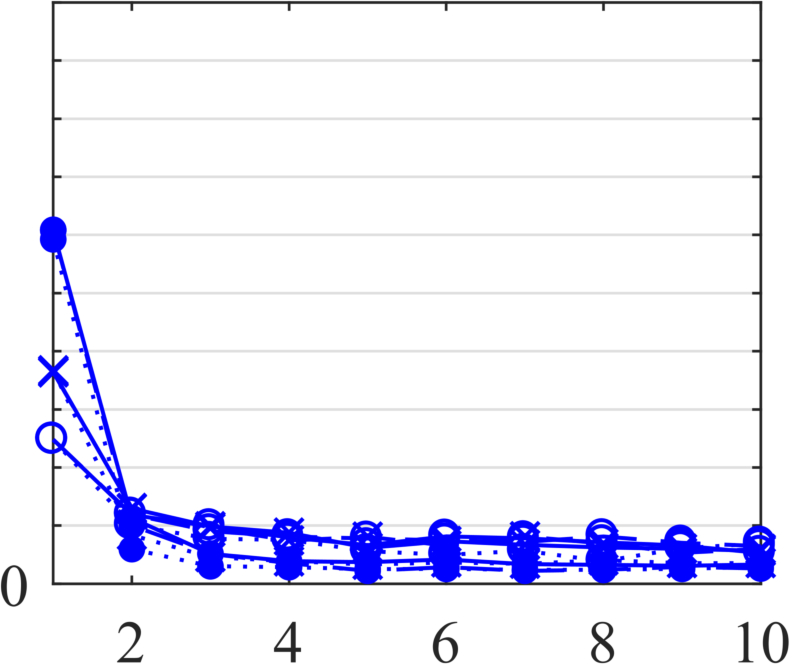}
	\end{subfigure}
	\caption{Evaluation of the metrics measuring the difference between previous and updated value of the $p_j^k$'s (see text for definitions) as a function of subproblems solved. Values obtained for the prostate (top) and lung (bottom) case when solved using the binary (red) and the fractional (blue) version of the algorithm. Note that the red and blue values are given by two different metrics and cannot be compared.}
	\label{fig:stopCriteria}
\end{figure}

\begin{figure}
	\centering
	\begin{subfigure}[b]{.7\textwidth}
		\flushleft\hskip7pt
		\hskip6pt\includegraphics[scale=.59]{legend20.png}\hskip20pt
		\hskip13pt\includegraphics[scale=.59]{legend11.png}\hskip20pt
		\hskip13pt\includegraphics[scale=.59]{legend7.png}
	\end{subfigure}
	\vskip-.25\baselineskip
	\begin{subfigure}[b]{.7\textwidth}
		\flushleft\hskip7pt
		\includegraphics[scale=.59]{legend240.png}\hskip20pt
		\includegraphics[scale=.59]{legend180.png}\hskip20pt
		\raisebox{-1.5pt}{\includegraphics[scale=.59]{legend120.png}}
	\end{subfigure}
	\vskip\baselineskip
	\begin{subfigure}[b]{.32\textwidth}
		\centering
		\includegraphics[scale=.75]{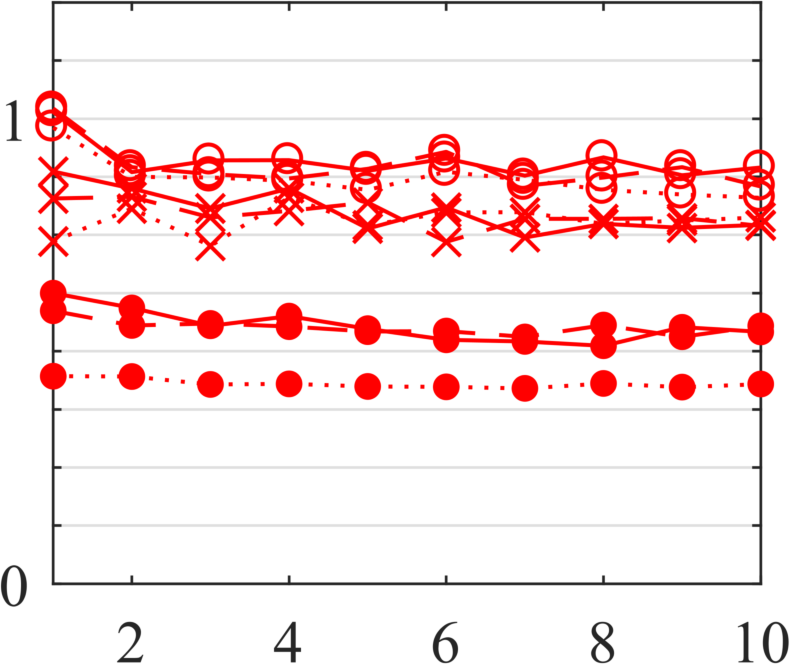}
	\end{subfigure}
	\hskip3pt
	\begin{subfigure}[b]{.32\textwidth}
		\centering
		\includegraphics[scale=.75]{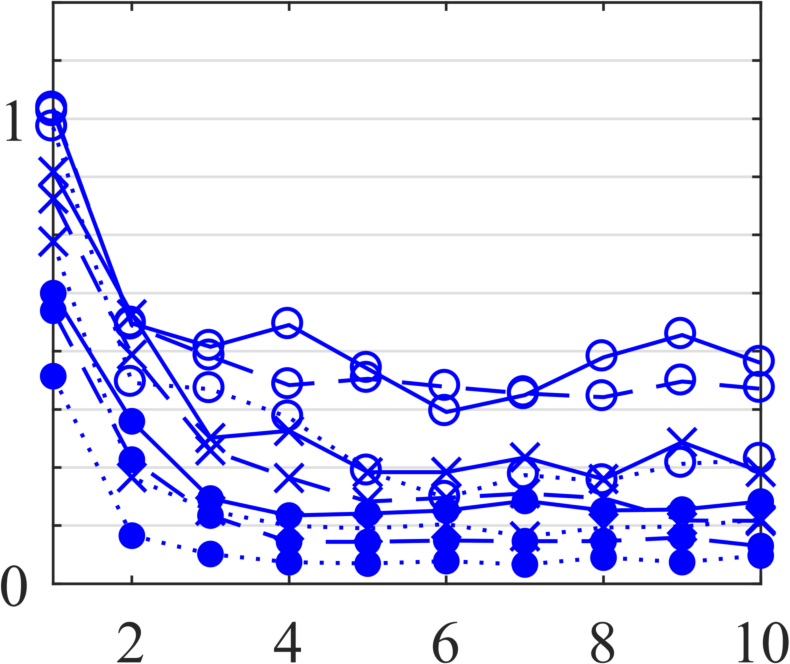}
	\end{subfigure}
	\vskip\baselineskip
	\begin{subfigure}[b]{.32\textwidth}
		\centering
		\includegraphics[scale=.75]{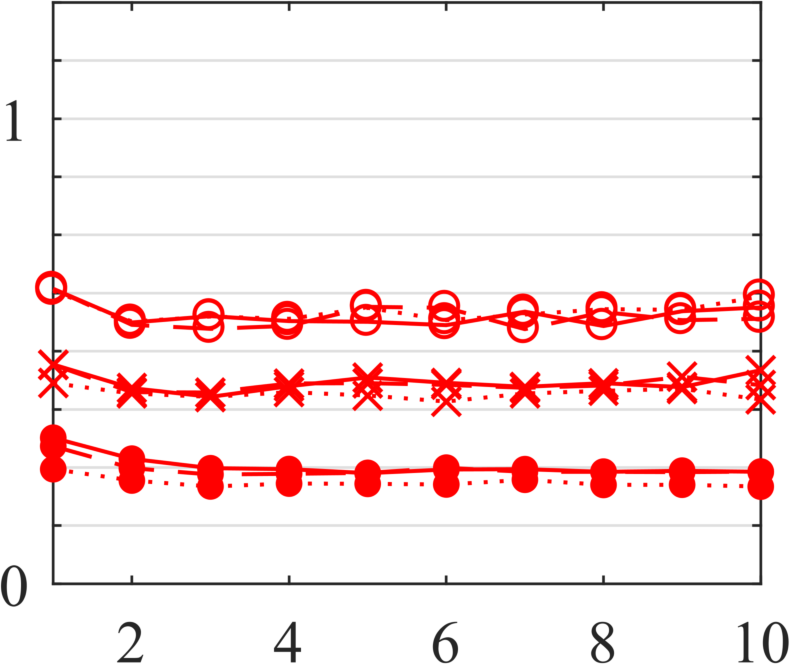}
	\end{subfigure}
	\hskip3pt
	\begin{subfigure}[b]{.32\textwidth}
		\centering
		\includegraphics[scale=.75]{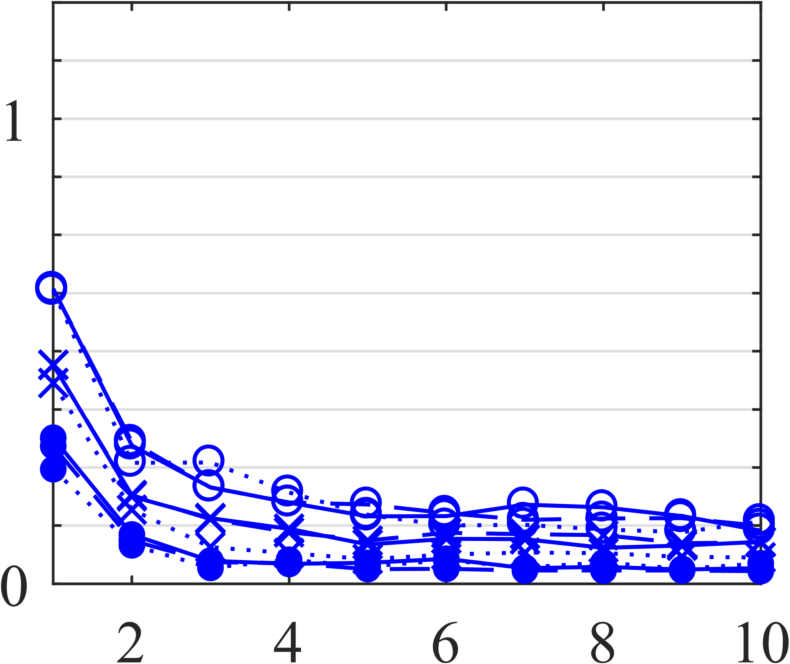}
	\end{subfigure}
	\caption{The (scaled) two-norm of the dose discrepancy between optimized and accurate dose as a function of subproblems solved. Results obtained for the prostate (top) and lung (bottom) case solved using the binary (red) and the fractional (blue) version of the algorithm.}
	\label{fig:discrepancy}
\end{figure}

To simulate the termination criteria used in \cite{papp2014} for the binary version of the algorithm, we use a metric that accumulates the absolute differences in index $k$ between the previous and updated control point for which $p_j^k = 1$ (thus, a quantity proportional to the angular difference between control points). For the fractional version, we use a similar metric that accumulates the absolute differences in previous and updated $p_j^k$ for all $k$. Evaluations of these two metrics are presented in Figures~\ref{fig:stopCriteria}. 
The two algorithm versions behave similarly: the metrics stagnate after a few iterations, which is in accordance with the observations in \cite{papp2014} where only two or three iterations were required. However, while lower values in the fractional-version metric is an indication of less deterioration in dose after accurate dose computation---a consequence of choosing the exact exposure fraction as fixed $p_j^k$---the same cannot be said about low values in the binary-version metric. An observation along these lines can be made in Figure~\ref{fig:discrepancy}, where the dose discrepancy between optimized and accurate dose is illustrated. The discrepancy obtained with the binary version is almost constant, whereas it decreases during the first few iterations of the fractional version of the algorithm according to a pattern similar to the termination metric in Figure~\ref{fig:stopCriteria}. The dependence of the discrepancy on the number of sweeps appears the strongest for the binary version, with the smallest discrepancy obtained for the plans with largest number of sweeps. 

\begin{figure}
	\centering
	\begin{subfigure}[b]{\textwidth}
		\flushleft
		\hskip6pt\includegraphics[scale=.59]{legend20.png}\hskip20pt
		\hskip13pt\includegraphics[scale=.59]{legend11.png}\hskip20pt
		\hskip13pt\includegraphics[scale=.59]{legend7.png}
	\end{subfigure}
	\vskip-.25\baselineskip
	\begin{subfigure}[b]{\textwidth}
		\flushleft
		\includegraphics[scale=.59]{legend240.png}\hskip20pt
		\includegraphics[scale=.59]{legend180.png}\hskip20pt
		\raisebox{-1.5pt}{\includegraphics[scale=.59]{legend120.png}}
	\end{subfigure}
	\vskip.5\baselineskip
	\begin{subfigure}[b]{.32\textwidth}
		\centering
		\includegraphics[scale=.75]{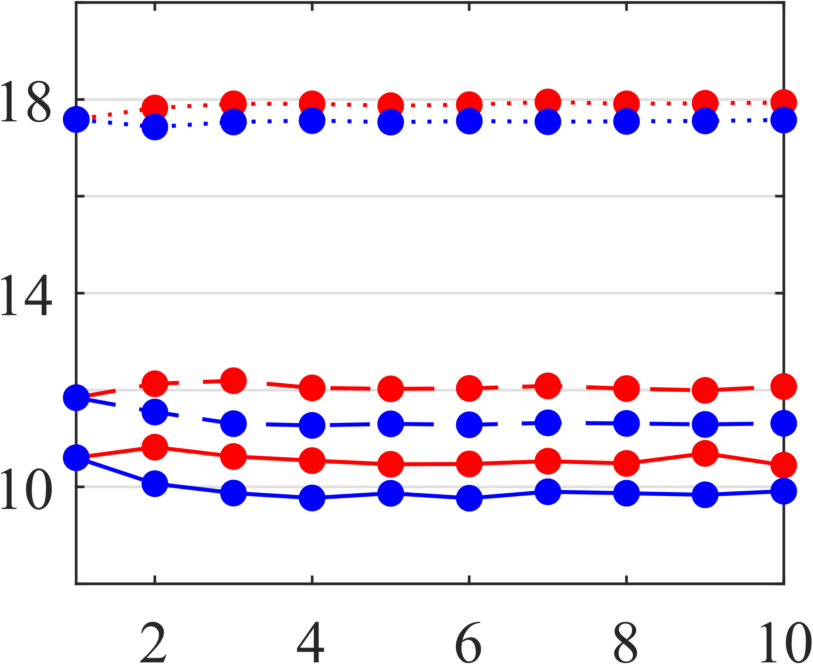}
	\end{subfigure}
	\hfill
	\begin{subfigure}[b]{.32\textwidth}
		\centering
		\includegraphics[scale=.75]{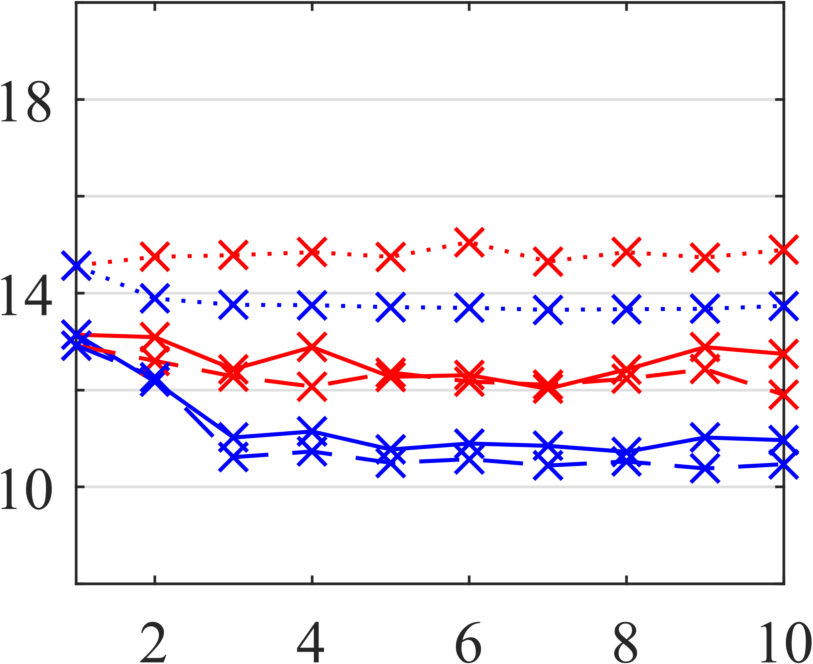}
	\end{subfigure}
	\hfill
	\begin{subfigure}[b]{.32\textwidth}
		\centering
		\includegraphics[scale=.75]{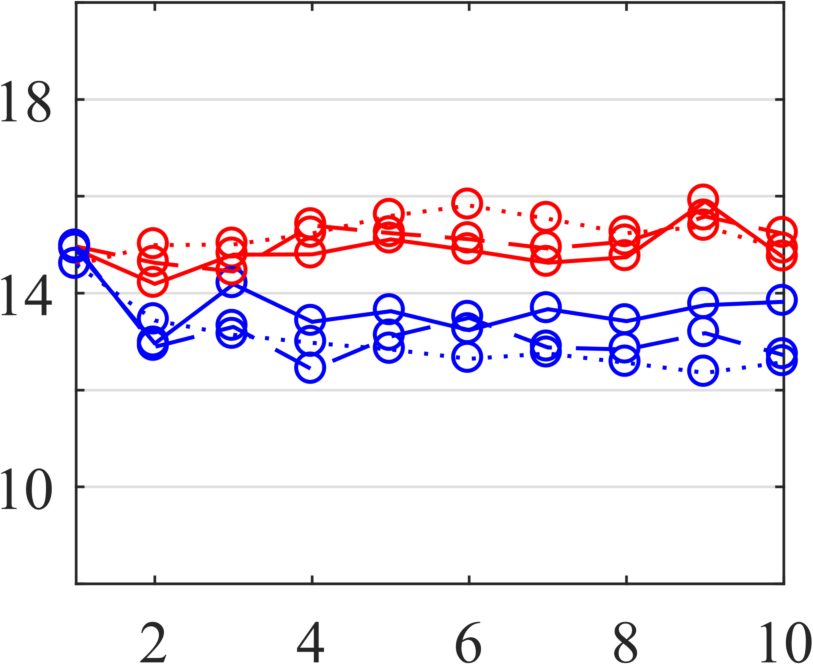}
	\end{subfigure}
	\vskip\baselineskip
	\begin{subfigure}[b]{.32\textwidth}
		\centering
		\includegraphics[scale=.75]{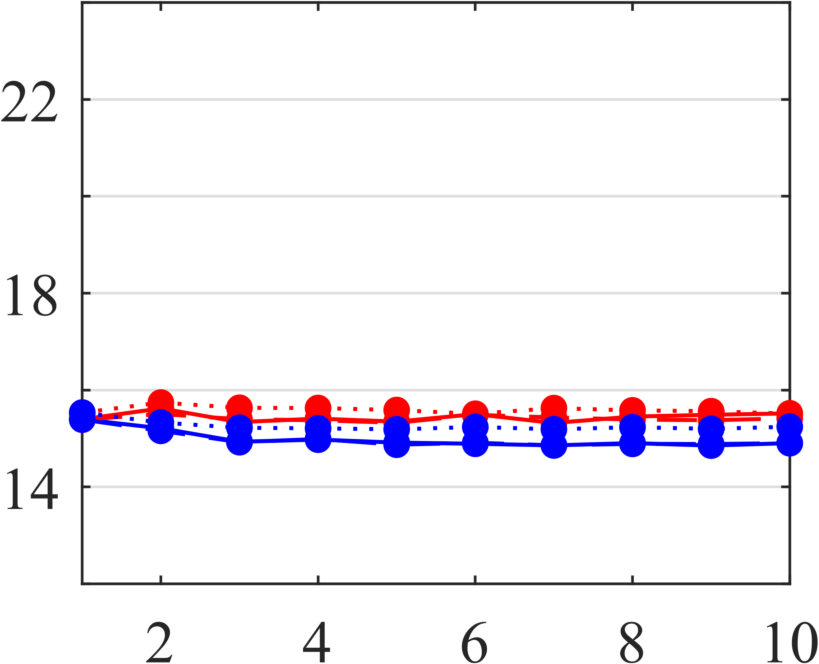}
		20 sweeps.
	\end{subfigure}
	\hfill
	\begin{subfigure}[b]{.32\textwidth}
		\centering
		\includegraphics[scale=.75]{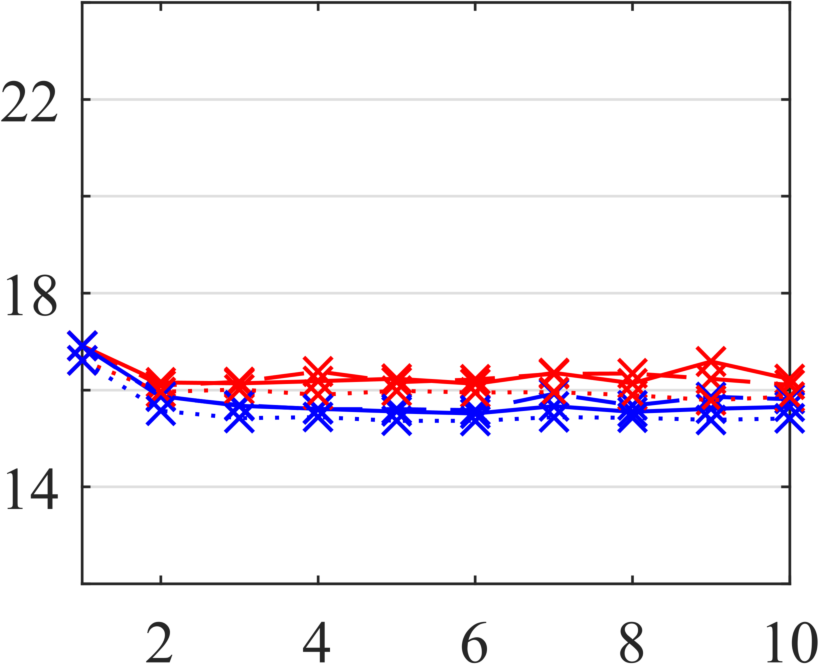}
		11 sweeps.
	\end{subfigure}
	\hfill
	\begin{subfigure}[b]{.32\textwidth}
		\centering
		\includegraphics[scale=.75]{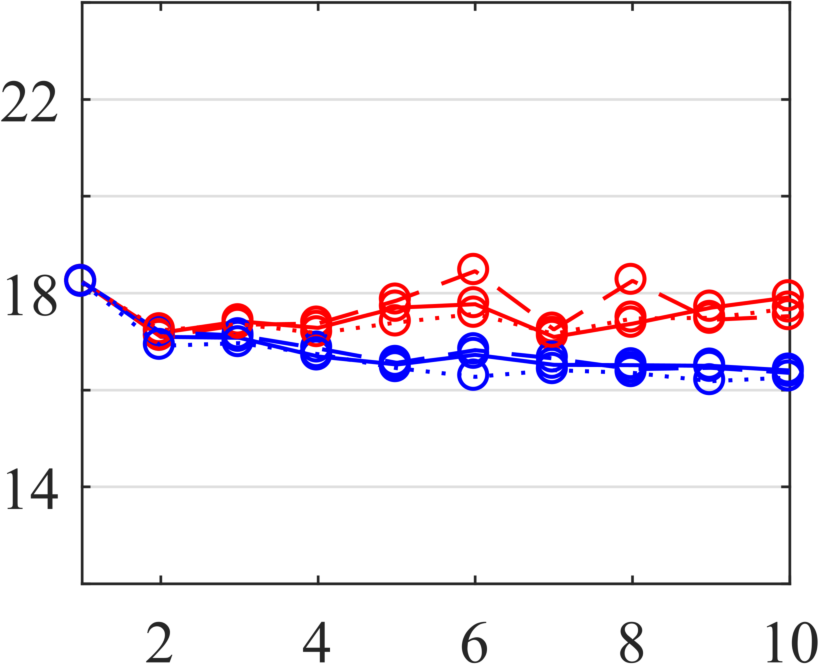}\\
		7 sweeps.
	\end{subfigure}	
	\caption{Objective function values evaluated for the accurate dose as a function of subproblems solved. Results obtained for the prostate (top) and lung (bottom) case solved using the binary (red) and the fractional (blue) version of the algorithm; results are subdivided into plots by number of sweeps.}
	\label{fig:objValSweeps}
\end{figure}

\begin{figure}
	\centering
	\begin{subfigure}[b]{\textwidth}
		\flushleft
		\hskip6pt\includegraphics[scale=.59]{legend20.png}\hskip20pt
		\hskip13pt\includegraphics[scale=.59]{legend11.png}\hskip20pt
		\hskip13pt\includegraphics[scale=.59]{legend7.png}
	\end{subfigure}
	\vskip-.25\baselineskip
	\begin{subfigure}[b]{\textwidth}
		\flushleft
		\includegraphics[scale=.59]{legend240.png}\hskip20pt
		\includegraphics[scale=.59]{legend180.png}\hskip20pt
		\raisebox{-1.5pt}{\includegraphics[scale=.59]{legend120.png}}
	\end{subfigure}
	\vskip.5\baselineskip
	\begin{subfigure}[b]{.32\textwidth}
		\centering
		\includegraphics[scale=.75]{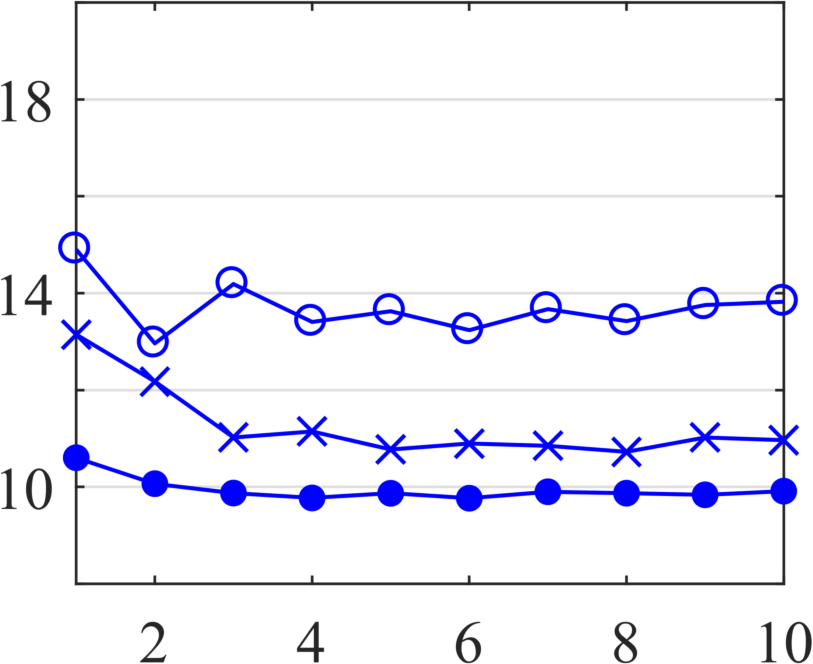}
	\end{subfigure}
	\hfill
	\begin{subfigure}[b]{.32\textwidth}
		\centering
		\includegraphics[scale=.75]{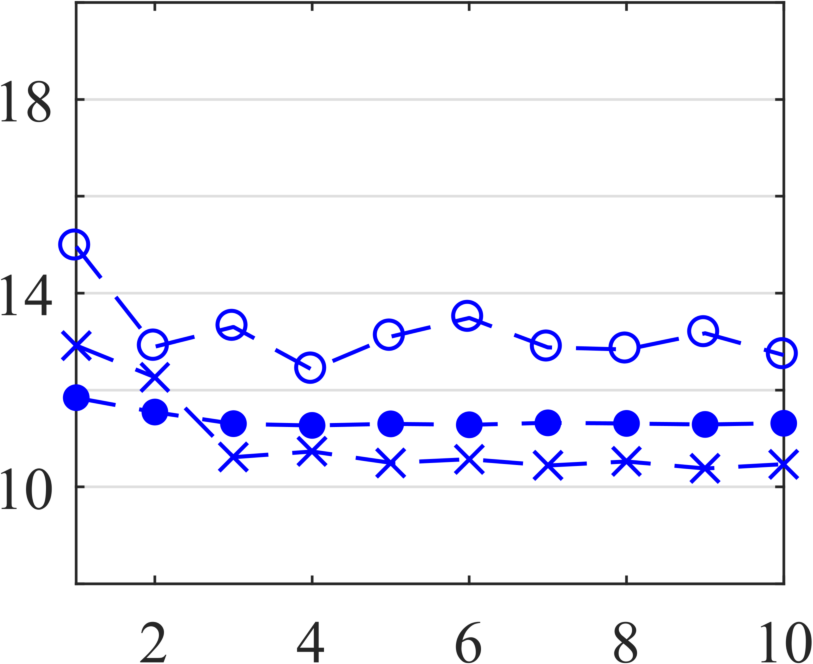}
	\end{subfigure}
	\hfill
	\begin{subfigure}[b]{.32\textwidth}
		\centering
		\includegraphics[scale=.75]{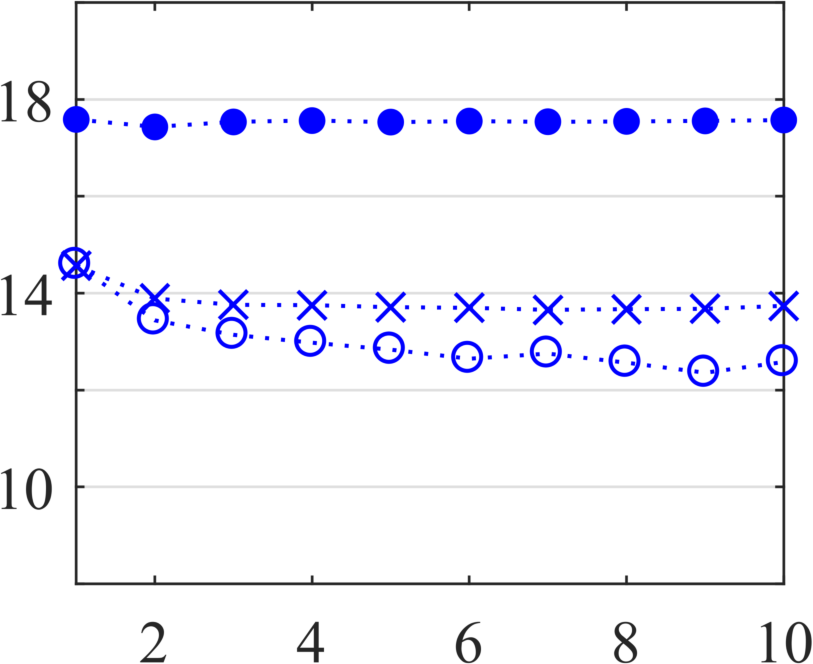}
	\end{subfigure}
	\vskip\baselineskip
	\begin{subfigure}[b]{.32\textwidth}
		\centering
		\includegraphics[scale=.75]{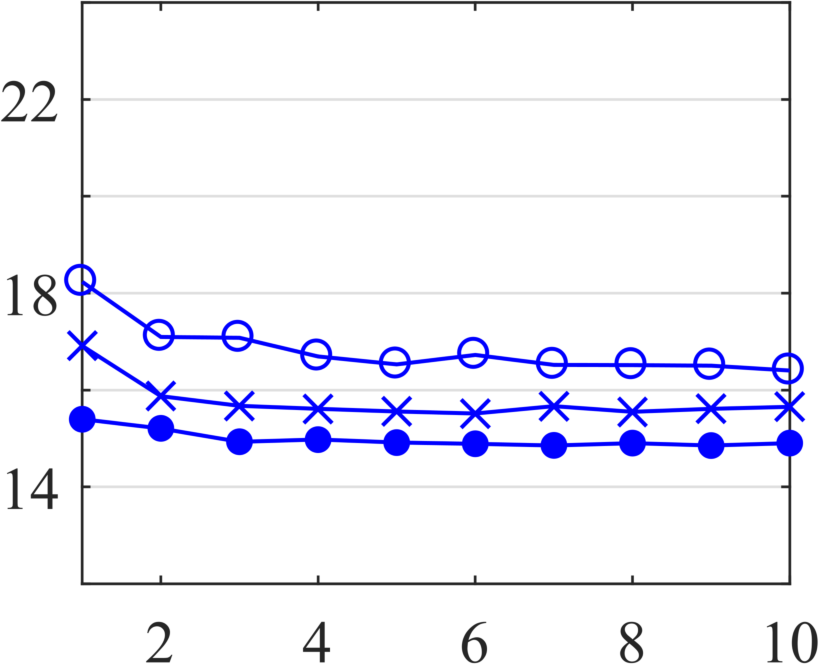}
		240 seconds.
	\end{subfigure}
	\hfill
	\begin{subfigure}[b]{.32\textwidth}
		\centering
		\includegraphics[scale=.75]{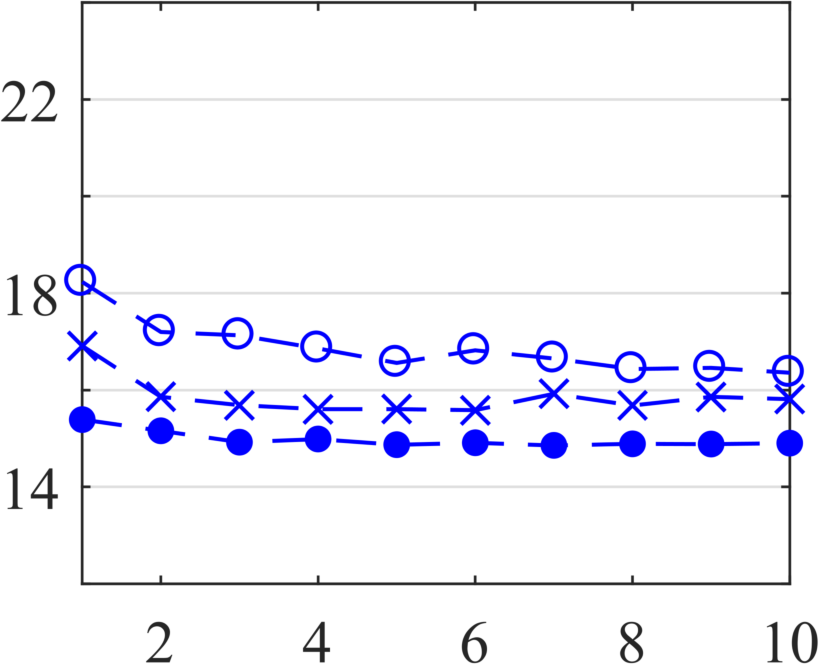}
		180 seconds.
	\end{subfigure}
	\hfill
	\begin{subfigure}[b]{.32\textwidth}
		\centering
		\includegraphics[scale=.75]{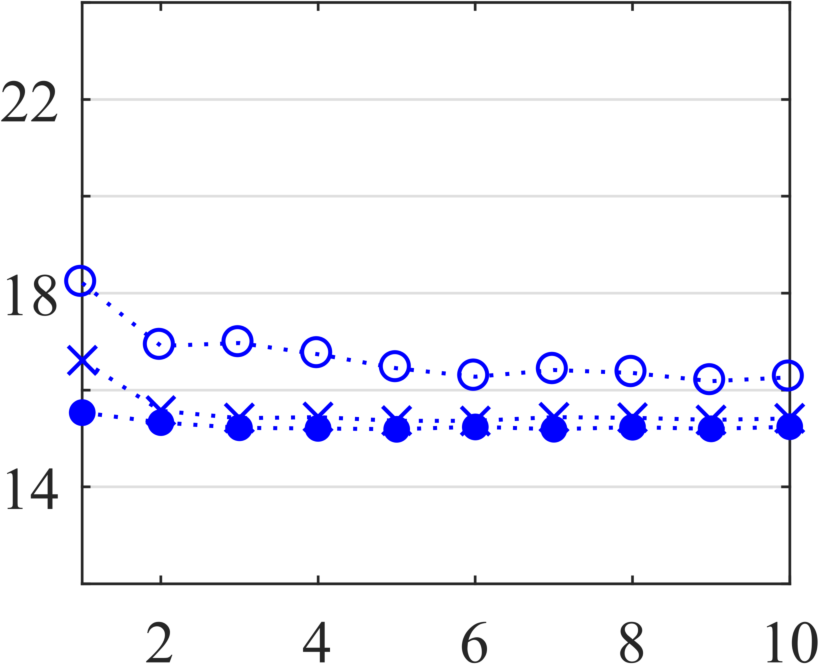}
		120 (150) seconds.
	\end{subfigure}	
	\caption{Objective function values from Figure~\ref{fig:objValSweeps}, here subdivided into plots by treatment time. Results obtained for the prostate (top) and lung (bottom) case when solved using the fractional version of the algorithm.}
	\label{fig:objValTimes}
\end{figure}

The plan quality in terms of objective function value evaluated for the accurate dose is presented in Figures~\ref{fig:objValSweeps} and~\ref{fig:objValTimes}. 
In Figure~\ref{fig:objValSweeps}, the plans are subdivided by number of sweeps. Comparing the two algorithm versions, a slight advantage in favor of the fractional version can be observed that, as expected, becomes more pronounced with fewer number of sweeps. In Figure~\ref{fig:objValTimes}, the values obtained when using the fractional version are rearranged, subdivided by treatment time. The effect of tighter time restrictions is the clearest in the prostate case where, e.g., the objective function values of the 20-sweep plan increase to eventually make this plan a less favorable alternative than both the 11- and 7-sweep plans. For instance, in this particular case, the 7-sweep plan is the best choice in terms of objective function value if a 120-second delivery is required. The variation in treatment time has less influence in the lung case. For such situations, the question of number of sweeps becomes more discrete: a decreasing treatment time will eventually become too tight for delivery of a large number of sweeps. For instance, as already mentioned, the 120-second restriction was relaxed to 150 seconds for all 20-sweep plans in order to fulfil the dose and delivery constraints. The results for the lung case still indicate that, e.g., delivering the 11-sweep plan in 120 seconds is nearly as good an option as delivering the 20-sweep plan in 150 seconds, with respect to objective function value.

\section{Discussion}\label{sec:Discussion}
A drawback with many heuristic methods is the lack of information about the distance to the global optimum. While the observations made in the numerical study indicate less deterioration in dose and better objective function values when applying our suggested fractional version of the Papp and Unkelbach algorithm~\cite{papp2014}, the heuristic nature of the algorithm makes it difficult to evaluate the improvement in proportion to the globally optimal plan. On the other hand, choosing the fractional version over the original binary version is not associated with any costs; their computational complexity, for instance, is identical. The decision to use the fractional version should therefore be uncontroversial and, as suggested by the results, the better alternative. 

In comparing plans delivered with different numbers of sweeps, we have chosen to report only objective function values. It should be mentioned that also the feasibility with respect to dose constraints has been evaluated, and that the fractional version of the algorithm resulted in plans of higher accuracy in this sense. However, a more clinical evaluation of the effects of fewer sweeps, e.g., using dose-volume histograms (DVHs), is not included in this study (though in our previous studies, we have observed good correlation with plan quality measures of the objective functions of \eqref{eq:propForm} \cite{engberg2017,engberg2018}). While DVHs give a more detailed view of the entire dose distribution, focus is easily placed on DVH features that are not controlled by any objectives or constraints. 

An arc-sequencing method has been suggested by Craft et al.~\cite{craft2012a} that successively decreases the number of sweeps, thus improves the delivery efficiency, until the fluence maps are no longer reproduced with sufficient precision. A drawback with methods that are focused on reproducing fluence maps is that, for longer arc segments, even perfectly reproduced fluence maps may give large differences in dose due to the larger variations in the dose deposition matrices. Similar to the Papp and Unkelbach algorithm, such methods thus rely on the final plan having relatively short arc segments and a large number of sweeps. In the present study, we have generated plans with only 11 and 7 arc segments. Results from the two patient cases indicate that even the 7-sweep plans could show the better objective function values, in case of a treatment time restriction approaching the typical delivery time of a regular VMAT plan (note that comparison has not been made to the plan quality of regular VMAT; however, recall that, e.g., regular VMAT is not as compatible with multicriteria optimization). Development of a method to, as in \cite{craft2012a}, dynamically determine the optimal---with respect to objective function value---number of sweeps given a certain delivery time restriction was beyond the scope of this study but is a possible direction of further research.

\section{Conclusion}
We have given an exact formulation of direct machine parameter optimization of sliding-window VMAT, by expressing the accurate dose as an explicit function of the sweeping leaf trajectories while taking into account the rotation of the gantry. The exact formulation is a nonsmooth optimization problem, and while to directly solve this formulation is not considered in this study, it has enabled us to generalize an algorithm previously suggested in the literature for generation of sliding-window VMAT plans. 

In the numerical study, plans have been generated with as few as 11 and 7 sliding-window sweeps, each delivered on a relatively large arc segment of the 360-degree arc. The purpose was to study the effects on the plan quality of a tight time restriction approaching the delivery time of regular (arbitrary leaf motion) VMAT. The results from the two patient cases show that, if requiring such an efficient sliding-window delivery, the few-sweep plans could give the better objective function values when compared to 20-sweep plans. While the plan quality naturally is not comparable to that obtained for 20-sweep plans with a generous time restriction, the few-sweep plans could be regarded as fast-delivery alternatives to static-gantry treatment plans for which 7-11 beam angles can be considered. The results furthermore show that our suggested version of the generalized algorithm performs better than the original algorithm in terms of better objective function value and less dose deterioration after accurate dose computation. These results are particularly pronounced for the plans with large arc segments.

\section*{Acknowledgement}
The authors thank Kjell Eriksson for valuable discussions. 

\bibliography{vmat}
\bibliographystyle{myplain} 

\appendix

\section{A MILP formulation of dose constraints}\label{app:MILP}
\allowdisplaybreaks
Every dose constraint of \eqref{eq:propForm} can be generalized into either an upper or lower bound on the voxel dose $d_i$. More precisely, upper bounds are obtained for the maximum dose and upper mean-tail-dose objectives/constraints, whereas the minimum dose and lower mean-tail-dose objectives/constraints imply lower bounds. To demonstrate the transformation into MILP constraints, it thus suffices to consider the two cases $d_i \leq \xi$ and $d_i \geq \xi$. For the upper-bound case, by the introduction of binary variables, we first obtain a nonlinear integer formulation:
\begin{align*}
	d_i \leq \xi \\
	\Leftrightarrow\enskip & \delta \sum_j \sum_{k \in K_b} P^{k\theta}_{ji}\, \max\big\{\min\big(l_j+\frac{\Delta}{2},\,t^{(k+\frac{1}{2})\theta}\big)\, - \\[-9pt]
	                       & \qquad\qquad\qquad\qquad\qquad\qquad\qquad\quad - \max\big(r_j+\frac{\Delta}{2},\,t^{(k-\frac{1}{2})\theta}\big),\, 0\, \big\} \leq \xi \\
	\Leftrightarrow\enskip & \delta \sum_j \sum_{k \in K_b} P^{k\theta}_{ji}\, \gamma_j^k \leq \xi, \\
						& \qquad\gamma_j^k \geq \min\big(l_j+\frac{\Delta}{2},\,t^{(k+\frac{1}{2})\theta}\big) - \max\big(r_j+\frac{\Delta}{2},\,t^{(k-\frac{1}{2})\theta}\big) \\
						& \qquad\gamma_j^k \geq 0 \\[5pt]
	\Leftrightarrow\enskip & \delta \sum_j \sum_{k \in K_b} P^{k\theta}_{ji}\, \gamma_j^k \leq \xi, \\
						& \qquad \gamma_j^k \geq b_1^{jk}\,(l_j-r_j), \\
						& \qquad \gamma_j^k \geq b_2^{jk}\,(l_j+\frac{\Delta}{2}-t^{(k-\frac{1}{2})\theta}), \\
						& \qquad \gamma_j^k \geq b_3^{jk}\,(t^{(k+\frac{1}{2})\theta}-r_j-\frac{\Delta}{2}), \\
						& \qquad \gamma_j^k \geq b_4^{jk}\,(t^{(k+\frac{1}{2})\theta}-t^{(k-\frac{1}{2})\theta}), \\
						& \qquad (\gamma_j^k \geq 0 \text{ implicit}) \\[5pt]
						& \qquad b_1^{jk}+b_2^{jk}+b_3^{jk}+b_4^{jk} = 1,\quad b_1^{jk},b_2^{jk},b_3^{jk},b_4^{jk} \;\text{ binary.}
\end{align*}
An equivalent MILP formulation can then be constructed by using the ``big $M$'' method, with which a nonlinear integer constraint such as
\[
	\gamma_j^k \geq b_1^{jk}\,(l_j-r_j)
\] 
is transformed using a sufficiently large $M$ into the linear integer constraint
\[
	\gamma_j^k \geq l_j-r_j - M(1-b_1^{jk}). 
\]
MILP formulations for the remaining three integer constraints are analogously constructed. For the lower-bound case, a ``big $M$'' MILP formulation is obtained directly:
\begin{align*}
	d_i \geq \xi \\
	\Leftrightarrow\enskip & \delta \sum_j \sum_k P^{k\theta}_{ji}\, \max\big\{\min\big(l_j+\frac{\Delta}{2},\,t^{(k+\frac{1}{2})\theta}\big)\, - \\[-9pt]
	                       & \qquad\qquad\qquad\qquad\qquad\qquad\qquad\quad - \max\big(r_j+\frac{\Delta}{2},\,t^{(k-\frac{1}{2})\theta}\big),\, 0\, \big\} \geq \xi \\
	\Leftrightarrow\enskip & \delta \sum_j \sum_k P^{k\theta}_{ji}\, \lambda_j^k \geq \xi, \\
						& \qquad\lambda_j^k \leq \max\big\{\min\big(l_j+\frac{\Delta}{2},\,t^{(k+\frac{1}{2})\theta}\big) - \max\big(r_j+\frac{\Delta}{2},\,t^{(k-\frac{1}{2})\theta}\big),\, 0\, \big\} \\[5pt]
	\Leftrightarrow\enskip & \delta \sum_j \sum_k P^{k\theta}_{ji}\, \lambda_j^k \geq \xi, \\
						& \qquad \lambda_j^k \leq l_j-r_j                                             + M b_5^{jk}, \\
						& \qquad \lambda_j^k \leq l_j+\frac{\Delta}{2}-t^{(k-\frac{1}{2})\theta}      + M b_5^{jk}, \\
						& \qquad \lambda_j^k \leq t^{(k+\frac{1}{2})\theta}-r_j-\frac{\Delta}{2}      + M b_5^{jk}, \\
						& \qquad \lambda_j^k \leq t^{(k+\frac{1}{2})\theta}-t^{(k-\frac{1}{2})\theta} + M b_5^{jk}, \qquad \lambda_j^k \leq M b_6^{jk}, \\[5pt]
						& \qquad b_5^{jk}+b_6^{jk} = 1,\quad b_5^{jk},b_6^{jk} \;\text{ binary.}
\end{align*}

\end{document}